\documentstyle[twocolumn,floats,aps]{revtex}
\newcommand{\be}{\begin{equation}}
\newcommand{\ee}{\end{equation}}
\newcommand{\bea}{\begin{eqnarray}}
\newcommand{\eea}{\end{eqnarray}}
\newcommand{\nl}{\vskip \baselineskip}
\begin{document}
\draft 
\twocolumn[\hsize\textwidth%
\columnwidth\hsize\csname@twocolumnfalse\endcsname
\title{\hbox{ \rightline{ \normalsize  UFRJ--IF--FPC/99}}
{Vortex Reconnection as the Dissipative Scattering of Dipoles}}
\author{L. Moriconi}
\address{Instituto de F\'\i sica, Universidade Federal do Rio de Janeiro,\\
C.P. 68528, Rio de Janeiro, RJ -- 21945-970, Brasil}
\maketitle
\begin{abstract}
We propose a phenomenological model of vortex tube reconnection at high 
Reynolds numbers. The basic picture is that squeezed vortex lines, formed by 
stretching in the region of closest approach between filaments, interact like dipoles (monopole-antimonopole pairs) of a confining electrostatic theory. The probability of dipole creation is found from a canonical ensemble spanned by foldings of the vortex tubes, with temperature parameter estimated from the typical energy variation taking place in the reconnection process. Vortex line reshuffling by viscous diffusion is described in terms of directional transitions of the dipoles. The model is used to fit with reasonable accuracy experimental data established long ago on the symmetric collision of vortex rings. We also study along similar lines the asymmetric case, related to the reconnection of non-parallel vortex tubes.
\end{abstract}
\vskip1pc] 
\narrowtext
There is growing evidence that the dynamics of vortex tubes is a necessary ingredient for a deeper understanding of turbulence. This view, initially suggested by images of the vorticity field produced through direct numerical simulations \cite{she1}, received strong support from the recent accurate determination of scaling exponents for the velocity structure functions, within a phenomenological theory which places filamentary configurations on a fundamental status \cite{she2}. However, the present knowledge on vortex dynamics is still far from being complete, so that even in simplified situations, as in the collision of vortex rings, a formal theory remains to be developed. This difficulty is related in part to the absence of comprehensive phenomenological descriptions that could provide a starting point for more elaborate discussions. 

We will focus our attention on the problem of vortex reconnection. Its importance -- considered more as an expectation in the long run -- relies on the idea that the global structure of turbulent flows may depend on topology changing processes, like the intertwining of closed vortex tubes. Previous theoretical studies on vortex reconnection \cite{ashurst,pumir,saffman,shelley} basically correspond to the case of low and moderate Reynolds numbers. These attempts may be regarded as the counterpart to interesting experimental observations reported by different groups on the collision of vortex rings \cite{fohl,oshima,schatzle}. On the other side, at higher Reynolds numbers, relevant effects come into play, as stretching and the possible existence of singularities \cite{chorin1,siggia}. Another important element to be considered at high Reynolds numbers is that vortex tube evolution is hardly reproducible, due to sensitivity to initial conditions, and one has to resort, therefore, to statistical methods. While it could seem there is no hope of an analytical treatment, since there are no standard techniques to find statistical measures in unstable dynamical systems, we show in this note that the well-known Boltzman distribution appears as a natural candidate from which plausible consequences may be derived.

In an experiment performed about 25 years ago, Fohl and Turner \cite{fohl}, 
studied the symmetric collision of two identical vortex rings in water at high 
Reynolds number ($ Re \sim 4000$), approaching each other with variable angle $2 \theta$. They found that the fusion of colliding vortex rings, both with radius $r$ and velocities $(0,v \sin \theta,v \cos \theta )$ and $(0, - v \sin \theta , v \cos \theta )$, into a single ring with velocity $\sim (0,0,-v/2)$ and radius $2r$ allways occurs in a first stage. The fused ring exhibits amplitude oscillations \cite{thomson,widnall}, so that a second stage characterized by a splitting reconnection takes place with probability $p=p(\theta)$. The two rings created after the second reconnection move in a plane perpendicular to the initial collision plane. An important feature in the experiment is the existence of a critical angle. For $\theta > \theta_c \simeq 16^\circ $, it holds $p(\theta)=1$, where as $p( \theta) \rightarrow 0$ as $\theta \rightarrow 0$. An explanation of $\theta_c$ was given by Fohl and Turner, based on the fact that the modes of amplitude oscillations, which describe perturbations around a vortex ring of radius $r$ and velocity $v$, have wavelength $\lambda_n = 2 \pi r/n$ and period
\be
T_n (r,v)={ {2 \pi r} \over {n(n^2-1)^{1 \over 2} v}} \ , \label{eq1}
\ee
with $n \geq 2$. In a collision defined by the angle $2 \theta$, the 
projection of the ring's velocity on the direction transverse to the symmetry plane is $v \sin \theta $. One may expect the amplitude of oscillations in the fused ring, which depends on the
collision angle, to be $2r$ when 
\be
v \sin \theta  = {{2r} \over {T_2(2r,v/2)} } \ . \ \label{eq2}
\ee
In this case, diametrically opposite parts of the fused ring will touch each 
other, producing reconnection. Eq. (\ref{eq2}) may be readily solved, yielding 
$\theta_c = \arcsin ( \sqrt 3 / 2 \pi ) \simeq 16^\circ$. 

There are important questions not answered on this problem. We would like to understand in a more detailed way the form of $p(\theta)$, taking into account its behavior at small angles. As we show below, this may be achieved in an effective  model where an important role is played by the dynamics of interacting dipoles.

To start, imagine two vortex tubes, locally antiparallel in some 
neighbourhood $\Omega$, both carrying the same flux and having identical 
circular cross sections. A description of the physical mechanism underlying 
reconnection was put forward by Saffman \cite{saffman}. In his model, the 
strain field shrinks $\Omega$ in the plane perpendicular to the vortex tubes, so that viscous annihilation of vortex lines occurs, reducing the circulation in
$\Omega$. Therefore, a pressure gradient along the vortex's cores develops, which increases strain and then viscous diffusion, enhancing reconnection in a self-induced way. The equations assumed to describe these steps give reasonable answers, in spite of some disagreements with real and numerical experiments \cite{saffman,shelley}. Saffman's model is in fact devoted to the situation of two antiparallel vortex tubes interacting at close enough distance. The model works better in the case of strong viscous diffusion (low Reynolds numbers), where vorticity amplification is not very large.

We suggest here a scenario of reconnection at high Reynolds numbers, when strecthing effects become relevant, which probably does not disagree with Saffman's model, since its application will be related to a different range of physical parameters. The picture we will consider is that in the collision of vortex tubes a system of ``dipoles", emerges after some stretching in a process characterized by very small energy lost. Reconnection is finished  with right-angle transitions and subsequent collapse of dipoles, through vortex line reshuffling by viscous diffusion. The main events are depicted in Fig. 1. It is important to keep in mind that dipoles are just effective structures which represent stretched vortex tube segments\footnote{We mean, throughout this work, an analogy with the {\underline{electric dipole}} definition.}.
In a more rigorous approach the reconnection problem should be addressed in terms of vortex sheets rather than quasi one-dimensional supports of vorticity, since numerical simulations \cite{pumir,anderson} show that vortex tubes flatten in the reconnection region due to stretching. Dipoles have to be regarded only as an useful approximation, from which it is possible to get phenomenological results in the simplest computational way.

We are interested in studying the energy of a flow given by two long vortex tubes, both carrying vorticity flux $\phi$, with stretched segments of length $\delta$ and separated by the distance $d$, as shown in Fig. 1. Vorticity may be decomposed as $\vec \omega = \vec \omega^{(1)} + \vec \omega^{(2)}$, where $\vec \omega^{(1)}$ is the field locally amplified by stretching, and $\vec \omega^{(2)}$ is the field at other places along the tubes. The energy $E=E(\delta,d)$ may be written as
\be
E=E_1+E_2+E_{12} \ , \ \label{eq3}
\ee
where, introducing the notation
 \be
\langle \vec \omega, \vec \eta \rangle \equiv
{1 \over {8 \pi}} \int d^3
\vec x d^3 \vec x' {1 \over {|\vec x - \vec x'|} }
  \omega_i (\vec x)
\eta_i (\vec x') \ , \ \label{eq4}
\ee
we have
\bea
&&E_1= \langle \vec \omega^{(1)} , \vec \omega^{(1)} \rangle
\ , \
E_2 =  \langle \vec \omega^{(2)} , \vec \omega^{(2)} \rangle
\ , \ \nonumber \\
&&E_{12} =  2 \langle \vec \omega^{(1)} , \vec \omega^{(2)} \rangle\ . \ \label{eq5}
\eea
To simplify expressions, the fator $1 / 8 \pi$ in (\ref{eq4}) will be supressed henceforth, corresponding to the replacement $\phi \rightarrow 2 \sqrt{2 \pi} \phi$. Taking the stretched segments to be identical, both with circular cross section of radius $\epsilon \ll \delta$, it follows that
\be
{E_1 \over {4 \phi^2 \delta}}=
 \ln( {2 \delta \over \epsilon})
-{\hbox{arcsinh}}( {\delta \over d } ) -1+ 
[({d \over \delta})^2 + 1]^{{1 \over 2}}
- {d \over \delta} \ . \
\label{eq6}
\ee
In the computation of $E_1$, the kernel in (\ref{eq4}) is regularized by means of
\be
{1 \over {|\vec x - \vec x'|}} \rightarrow {1 \over {[(\vec x - \vec x')^2 + \epsilon^2]^{1 \over 2}}} \ . \ \label{eq7}
\ee
The contribution to the energy which comes from the interaction between the stretched and non-stretched parts is
\bea
E_{12} &&= \langle \vec \omega^{(1L)} , \vec \omega^{(2L)} \rangle +
\langle \vec \omega^{(1R)} , \vec \omega^{(2R)} \rangle  \nonumber \\
&&+ \langle \vec \omega^{(1L)} , \vec \omega^{(2R)} \rangle +
\langle \vec \omega^{(1R)} , \vec \omega^{(2L)} \rangle \ , \ \label{eq8}
\eea
with the superscripts L and R denoting the left and right vortex tubes. The 
effect of screening between the two vortex tubes is evaluated from the 
following estimative
\be
{E_{12} \over { 4 \phi^2  \delta }} \sim 
\sum_{n=1}^\infty \left [ {1 \over n}
- {\delta \over {[n^2 \delta^2+d^2]^{ {1 \over 2} } } } \right ] \ . \ \label{eq9}
\ee
The above expression is derived from a discretized version of the integral (\ref{eq4}) in terms of vortex tube segments of length $\delta$, as discussed in ref. \cite{chorin2}, where we additionally used reflection symmetry (invariance of the vorticity field under the interchange L $\leftrightarrow$ R. In the reconnection experiments, $\delta$ typically fluctuates around some mean value $\bar \delta  > d$. This inequality implies, with (\ref{eq6}) and (\ref{eq9}), that for $\epsilon / \bar \delta \ll 1$ the interaction energy $\bar E_{12}$ may be neglected, when compared to $\bar E_{1}$, and, therefore, $E \simeq E_1 + E_2$ (as an example, consider the choice $\epsilon / \bar \delta \sim 10^{-2}$, which gives $E_{12}(\bar \delta, d)/E_1 (\bar \delta, d) < 0.1$). We conclude that the positive energy variation due to stretching is local, being compensated by reduction of energy through foldings of the vortex tubes \cite{chorin2}.

There is a close relationship between the stretched vortex tube segments
and dipoles of a confining electrostatic theory. Let $\vec p_1 (\vec x)$ and $\vec p_2 (\vec x)$ be the dipole moments of two charge distributions, respectively defined in compact regions $\Omega_1$ and $\Omega_2$. The interaction energy associated to a linear confining potential is
\be
E_{int} \sim \int d^3 \vec x d^3 \vec x' |\vec x-\vec x'| \vec \nabla\cdot \vec p_{1}
(\vec x)
\vec \nabla \cdot \vec p_{2}(\vec x') \ . \ \label{eq12}
\ee
Through integration by parts we get
\be
E_{int} \sim \langle \vec p_1 , \vec p_2 \rangle
- (\vec p_1 (\vec x) \cdot \vec r)
(\vec p_2 (\vec x') \cdot \vec r) |\vec r|^{-3}
\ , \ \label{eq13}
\ee
where $\vec r =\vec x - \vec x'$. Taking $\Omega_1$ and $\Omega_2$ as 
oriented segments, parallel to the vorticity field and parametrized by $(0,0,s)$ and $(-d,0,\delta - s)$, with $0 \leq s \leq \delta$ and $\delta \ll d$, the scalar products in the second term in eq. (\ref{eq13}) may be neglected. In this situation the result may be identified with the interaction energy associated to vorticity fields given by $\vec p_1$ and $\vec p_2$.

An alternative and straightforward way to obtain the connection with 
confining electrostatics is to represent the dipoles as ``monopole-antimonopole" pairs. This is done by replacing the stretched vortex tube segments by monopoles at positions $(0,0,\delta)$ and $(-d,0,0)$ and antimonopoles at $(0,0,0)$ and $(-d,0,\delta)$. These points are just the boundaries of $\Omega_1$ and $\Omega_2$. The monopole (antimonopole) is the source (sink) of a radially symmetric vorticity field, which is not solenoidal -- and hence deprived of direct physical meaning. The field of a monopole-antimonopole pair is given as $\vec \omega = \vec \omega^{(+)} + \vec \omega^{(-)}$, where
\be
\omega^{(\pm)}_i (\vec x)= \mp {\phi \over {4 \pi}} \partial_i
{1 \over {|\vec x - \vec x_\pm|} } \ . \ \label{eq14}
\ee
Above, $\vec x_\pm$ gives the position of the monopole with charge 
$\pm \phi$. To compute the energy of a system of monopole-antimonopole pairs, it is necessary to regularize infrared divergencies. Defining the flow inside a large sphere of radius $\Lambda \rightarrow \infty$, we will have
\be
\langle \vec \omega^{(p)} , 
\vec \omega^{(q)} \rangle = 2 \eta (p,q) \phi^2
(\Lambda - |\vec x_p - \vec x_q|) \ , \ \label{eq15}
\ee
where $p,q = \pm$, and $\eta(p,q)=1$ for $p=q$, otherwise $\eta(p,q)=-1$.
The energy of the interacting monopole-antimonopole pairs considered here
is, then\footnote{Note that the infrared divergencies cancel for a neutral system of monopoles.},
\be
E = 4 \phi^2 [ \delta +d-(d^2 + \delta^2)^{{1 \over 2}} ]\ . \ \label{eq16}
\ee
When $\delta \ll d$, the above expression differs from eq. (\ref{eq6}), only by self-energy terms, which are independent of $d$ (for $d/ \delta >1$ the agreement is better than $90 \%$).
Since the interaction potential is linear, the force between monopoles has constant strength and is directed along the line joining the particles. This suggests, in the description of stretched vortex tube segments as monopole-antimonopole pairs, that whenever $\delta > d$, reconnection takes place. This is in fact the same condition that follows from $E_1( \delta, d) > E_1(d, \delta)$, using the more precise result, eq. (\ref{eq6}). These are the energies for the configurations shown in Fig. 1. Furthermore, the energy dissipated in the reconnection process is $\Delta E = E_1( \delta,d) - E_1(d, \delta)$.

To model the symmetric collision of vortex rings, it is necessary to know how $d$ depends on the collision angle. This may be obtained by replacing the numerator $2r$ in eq. (\ref{eq2}) by $2r-d/(2c_1)$, which we propose to be the oscillation amplitude at angle $\theta$, where $c_1$ is a phenomenological parameter. We find $d (\theta) = 4 c_1 r( 1- 2 \pi \sin \theta / \sqrt 3  )$. As only small angles are involved ($0 \leq \theta \leq \theta_c \simeq 16^\circ$), this expression may be effectively thought as a linear interpolation between the maximum value $d \equiv 4 c_1 r$, which occurs at $\theta = 0^\circ$, and the minimum value $d=0$, at $\theta = \theta_c$. Another important parameter is the mean length of the stretched vortex tube segments. We define it as $\bar \delta= c_2r$. It is convenient to use dimensionless units where $ \bar \delta= 1$ and
\be
d (\theta) = 4 {c_1 \over c_2} (1- {2 \pi \over \sqrt 3} \sin \theta )
\ . \ \label{eq17}
\ee
Due to the additive property of energies, $E=E_1+E_2$,
we may interpret the random behavior of $\delta$ as being related to
fluctuations of $E_1$ in a canonical ensemble. The elements of the canonical ensemble correspond to folded configurations of the non-stretched parts of the vortex tubes, which act like a reservoir exchanging energy with the stretched region. The probability density to have stretching length $0 \leq \delta \leq \infty$ is
\be
\rho ( \delta,d) = Z^{-1}
\exp [ - \beta E_1(\delta,d) ]\ , \ \label{eq18}
\ee
where $\beta$ is the ``inverse temperature" and
\be
Z = \int_0^\infty dx \exp [ - \beta E_1(x,d)] \ . \ \label{eq19}
\ee
is the partition function. The definition $\bar \delta =1$ is used to find the temperature, $\beta^{-1} \simeq E_1 (1.0,d)$, which is the energy necessary for the creation of stretched segments of length $\bar \delta$. 
The probability to have reconnection is, thus,
\be
p(\theta) = \int_{d (\theta)} ^\infty dx \rho (x,d(\theta)) \ . \ \label{eq20}
\ee
It is not our purpose to derive the above statistical mechanical correspondence from first principles; we take it as a hypothesis. The main problem would be to show that energy fluctuations take place in a time scale much larger than the one of reconnection ($\sim \epsilon^2 / \phi$, see \cite{ashurst}). On the other hand, the relaxation time for the vortex system to reach thermal equilibrium has to be much smaller than the time spent for the whole process ($\sim r^2 / \phi$), lasting from the fusion of the vortex rings up to the instant of split. Amplitude oscillations may be an important aspect of such an analysis: at large $n$, the wave velocity is $\lambda_n/ T_n \sim n v$, according to eq. (\ref{eq1}). Therefore, ``thermal equilibrium" would be assured by the fast propagation of perturbations along the vortex tubes. Actually, this equilibrium is not stable, due to the attraction between dipoles. It is worth noting that a more complete study of the right-angle transition of dipoles would probably deal with methods of non-equilibrium statistical mechanics. 

We need to set the value of the dimensionless phenomenological parameter $c_1/c_2$. This is done in principle by searching for the best agreement with experimental data, but we do not expect this parameter to significantly depart from unit. The reason is that $\bar \delta \sim r$, as indicated in numerical and real experiments, and $d(0^\circ)\sim 4r$, if one assumes that the amplitude of oscillations in the fused ring vanishes when $\theta=0^\circ$. We note that $d(0^\circ) / \bar \delta = 4 c_1/c_2$ is a quantity suitable of experimental determination. The resulting $p(\theta)$ is shown in Fig. 2, with $c_1/c_2 =1.0$ and $\epsilon=0.01$. As a matter of fact, the form of $p(\theta)$ is not altered in an important way for $\epsilon < 0.1$. In the limit $\epsilon / \bar \delta \rightarrow 0$, we may solve eq. (\ref{eq20}) exactly, to get that the probability of reconnection decreases exponentially with the distance between the vortex tubes, that is, $p(\theta) = \exp[ - d(\theta)/ \bar \delta]$.

We may proceed through similar computations to study the asymmetric scattering of vortex rings, aiming at predictions that could be tested in future experiments.

In the symmetric case considered so far, let $A$ be the point where
the colliding vortex rings first touch. $A$ is diametrically opposite to points $B$ and $C$, which are placed in different rings. The angle at the vertex $A$, defined by the segments $AB$ and $AC$ is $180^\circ - 2 \theta$. We study the asymmetric collision obtained from the following initial configuration: while the ring which contains $C$ is fixed, the other ring is rotated around the axis $AB$ by the angle $\alpha$. This is precisely the angle between the vortex tubes at the point of contact.

We want to find now the probability  $p(\theta,\alpha)$ for the splitting reconnection. As the fused ring evolves, points $B$ and $C$ move toward each other with relative velocity
\be
2v \sin \theta \cos^2 (\alpha / 2) \ . \ \label{eq21}
\ee
This amounts to replacing $ \sin \theta $ which appears in eq. (\ref{eq17}) by $\sin \theta \cos^2 ( \alpha / 2)$, to define $d(\theta,\alpha)$, the distance between the non-parallel dipoles. Assuming that the second reconnection occurs close to points $B$ and $C$, with stretched vortex tubes keeping the initial relative angle $\alpha$, all we need to know is the energy of such a configuration. Neglecting the interaction terms in the expression for the energy, which depend on the distance between dipoles, the condition for reconnection is then $E_1(\delta) > E_1(\ell(\theta,\alpha))$, where
\be
\ell(\theta,\alpha)=[d^2(\theta,\alpha)+\delta^2 \sin^2 (\alpha / 2)]^{1 \over 2} \ . \ \label{eq22}
\ee
Therefore, reconnection occurs when $\delta > \ell ( \theta, \alpha)$, or, equivalently, 
$\delta > d( \theta, \alpha) / \cos (\alpha /2)$. We just have to replace the lower bound $d(\theta)$ in the integral (\ref{eq20}) by $d( \theta, \alpha) / \cos (\alpha /2)$. We carried out computations using the same set of parameters $\beta$ and $c_1/c_2$ for the former case ($\alpha = 0$). It is possible that deviations grow as $\alpha$ gets larger, where the dipole model may loose its applicability. However, there is some indication that reconnection is supressed at large $\alpha$ \cite{koplik}, which is also verified through an explicit computation of $p(\theta,\alpha)$. In Fig. 3 $p(\theta, \alpha)$ is shown with $0 \leq \alpha \leq 180^\circ$, for $\theta = 15^\circ$ and $\theta=17^\circ$. The latter situation is perhaps the more interesting, because of the plateau given by $p(\theta, \alpha)=1$ at small values of $\alpha$.

To summarize, we studied the problem of vortex reconnection at high Reynolds numbers, where stretching effects become important. A simple correspondence with confining eletrostatics and statistical mechanics allowed us to investigate the way how ``dipoles", i.e., stretched vortex tube segments, behave in the process of reconnection. The initial dipole configuration evolves, thanks to diffusion, towards a state which links the interacting vortex tubes. The measurements of Fohl and Turner \cite{fohl} for the probability of a splitting reconnection after the initial merger of vortex rings are successfully accounted by the present model. We also defined some predictions concerning the case of asymmetric collisions, to be compared with possible future experimental observations.
\nl

{\leftline {\bf Acknowledgments}}
\nl

This work was partially supported by CNPq.

\newpage
\centerline{{\bf{FIGURE CAPTIONS}}}
\nl
{\bf{Fig. 1}}
\nl
Sketch of the configuration of the vortex tubes in the region of closest approximation. An attractive force betwen the oppositely oriented stretched segments (``dipoles'') is followed by vorticity cancelation (not represented in the picture) and then by the right-angle transition of the dipoles.
\nl
\nl
{\bf{Fig. 2}}
\nl
The probability $p(\theta)$ for the occurrence of a splitting reconnection.
Dots represent the values measured by Fohl and Turner. The continuous line
is the prediction of the dipole model obtained with $\beta = 1/E_1(1.0,d)$,
$c_1/c_2 =1.0$, and $\epsilon=0.01$.
\nl
\nl
{\bf{Fig. 3}}
\nl
The probability $p(\theta,\alpha)$ for the ocurrence of a splitting reconnection in the asymmetric case, where $\theta$ is fixed and $0\leq \alpha \leq 180^\circ$. a) $\theta=15^\circ$.
b) $\theta=17^\circ$.

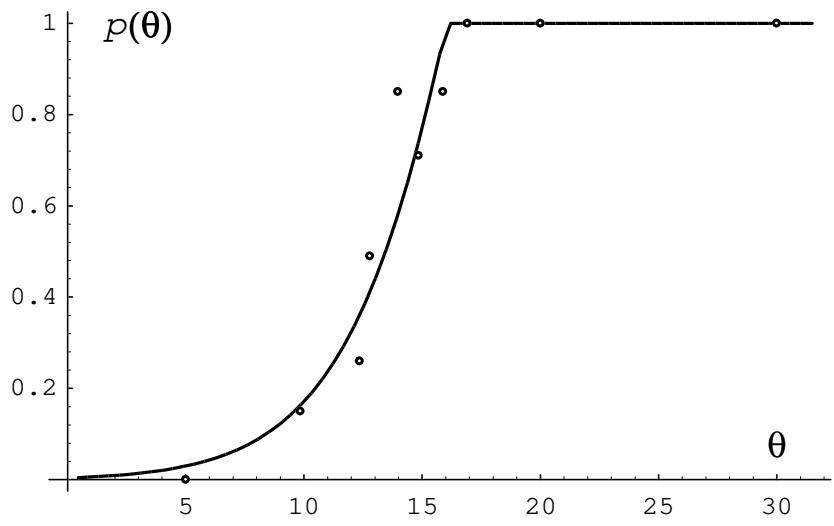
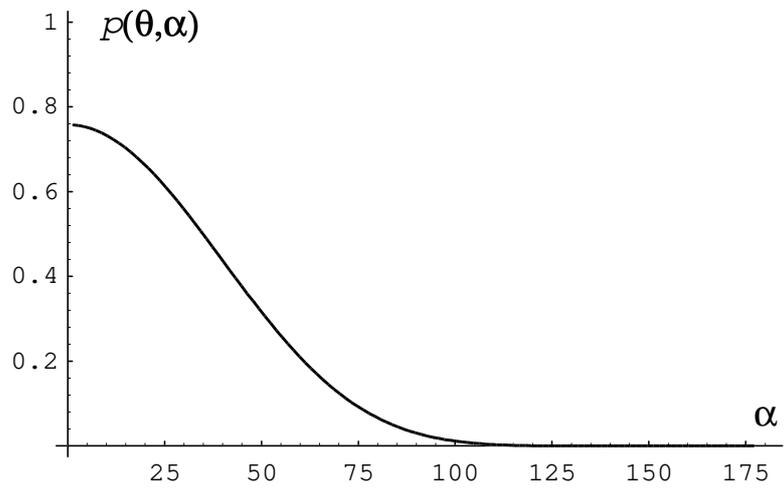
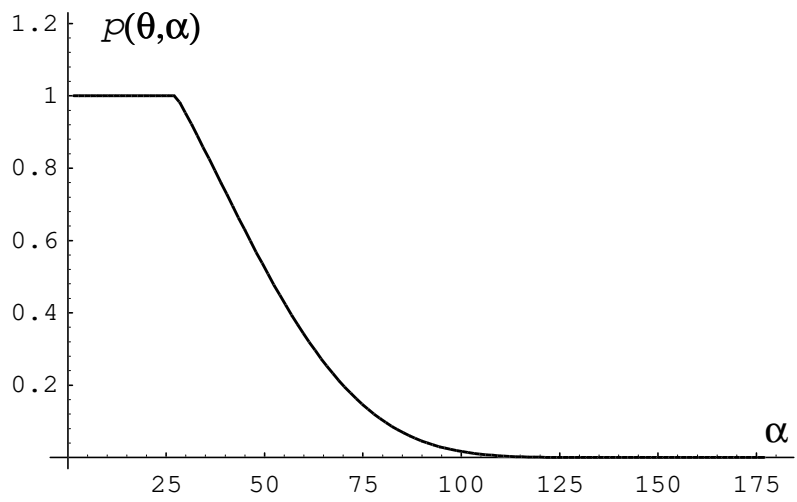
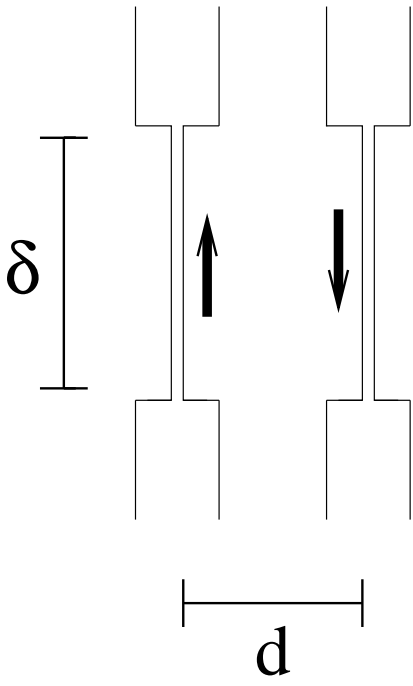
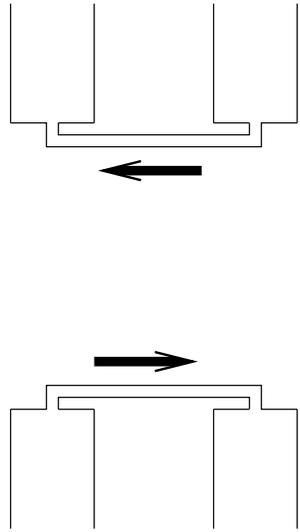
\end{document}